\documentclass[sigconf,nonacm]{acmart}
\begin{document}
\title[Touching Projected Agents through the DAT Screen]%
{Touching Projected Agents through the DAT Screen}
\subtitle{System Design and Questions of Interaction Boundaries}
\author{Takaya Tamamoto}
 \affiliation{\institution{Kansai University}\city{Takatsuki}\country{Japan}}
 \author{Tomoko Yonezawa}
 \affiliation{\institution{Kansai University}\city{Takatsuki}\country{Japan}}
 \renewcommand{\shortauthors}{Tamamoto and Yonezawa}
\begin{abstract}

Touching a virtual agent involves more than detecting a finger on a display: the contact surface can deform and resist the user's push, and the agent can respond to that action. We describe the Desktop Amid Touch (DAT) screen, which combines spatial projection onto a semi-transparent, deformable mesh with touch sensing and motorized control of resistance to indentation. 

An initial prototype detected a single touch and estimated its position and depth from tension changes around a flexible projection screen. A subsequent prototype added motorized tension control to vary the screen's resistance to indentation. We describe the tension-controlled DAT implementation and its demonstrations, distinguishing what has been observed from what remains an open design question. 

These capabilities raise a design question: should touchable virtual agents respond as though they had preferences or boundaries, and under what conditions should such behavior be implemented on systems like DAT. The present paper describes a system and frames this question; it does not establish how users interpret an agent's refusal or consent.

 \end{abstract}

\begin{teaserfigure}
  \centering
  \includegraphics[width=\textwidth]{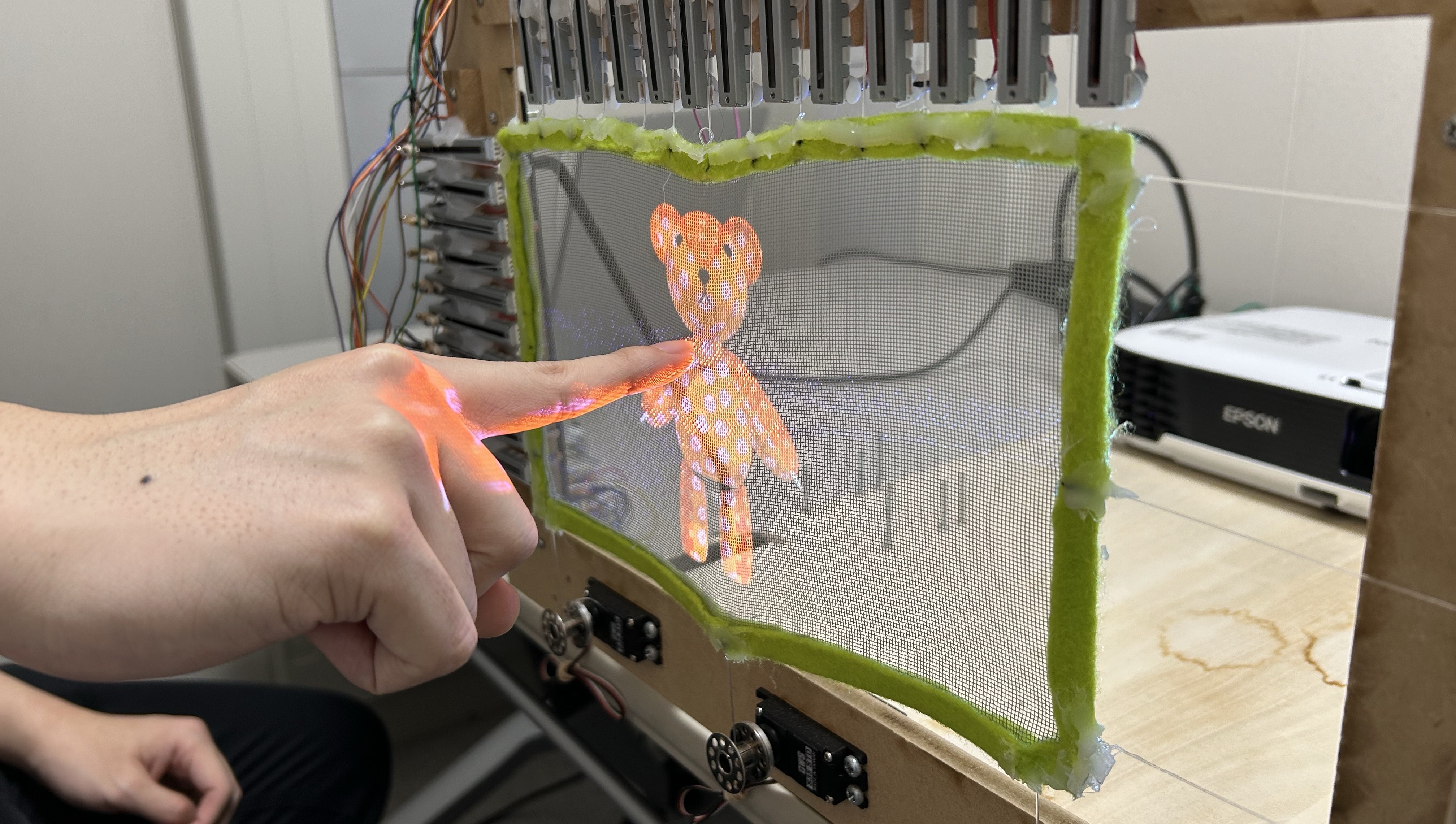}
  \caption{A user reaching toward an agent projected on the DAT screen.}
  \Description{A projected agent appears on a semi-transparent mesh
    as a user's finger approaches it.}
  \label{fig:teaser}
  \begin{minipage}{\textwidth}
  \footnotesize
  \centering
  \emph{Model credit:} \emph{Usagi to Kuma no Nuigurumi}
  by atuyu (@atuyu\_no01), modified by the authors
  (additional eye variants);
  \url{https://bowlroll.net/file/146474}.
  \end{minipage}
\end{teaserfigure}

\maketitle

\begin{figure*}[t]
  \centering
  \begin{minipage}[b]{0.30\textwidth}
    \centering\includegraphics[width=\linewidth]{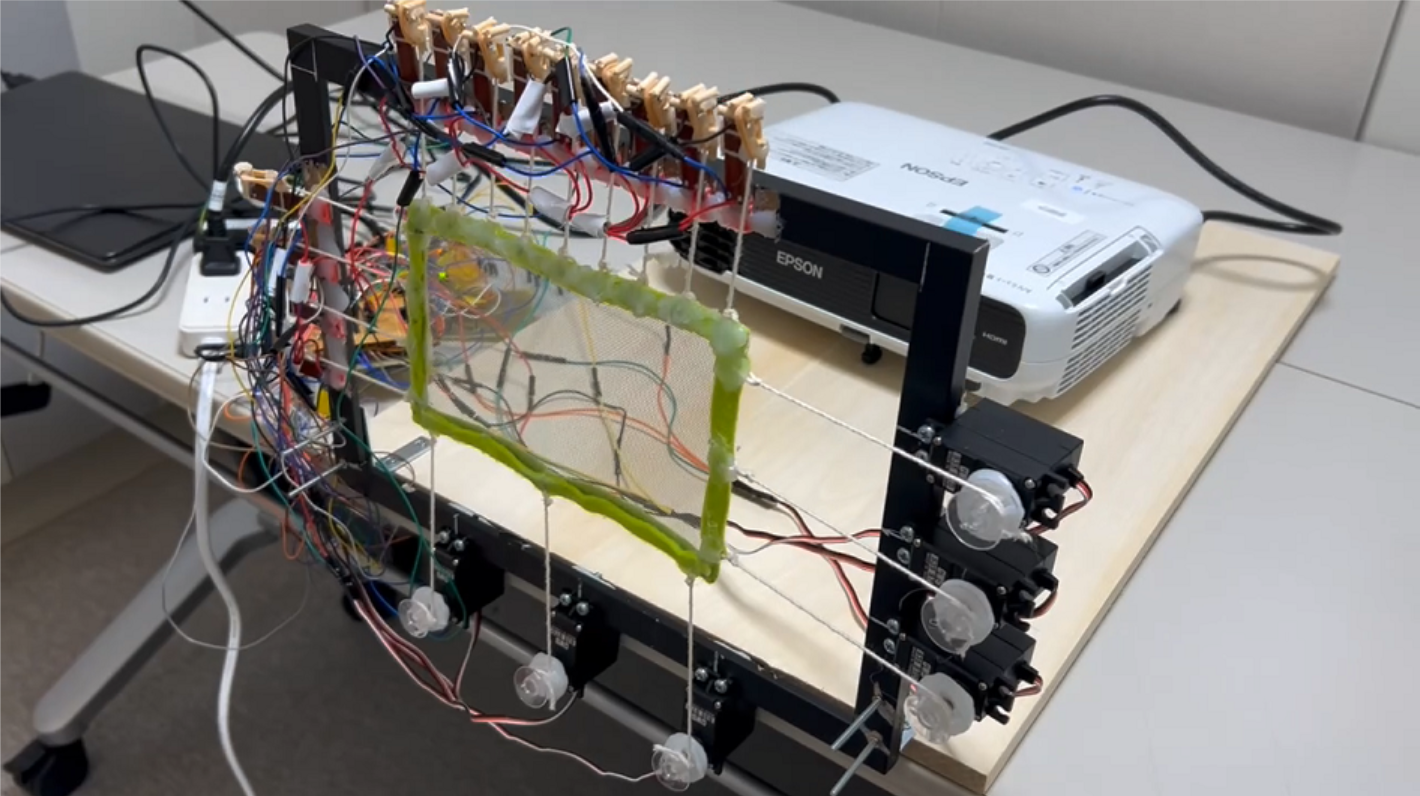}\\
    (a) System
  \end{minipage}\hfill
  \begin{minipage}[b]{0.47\textwidth}
    \centering\includegraphics[width=\linewidth]{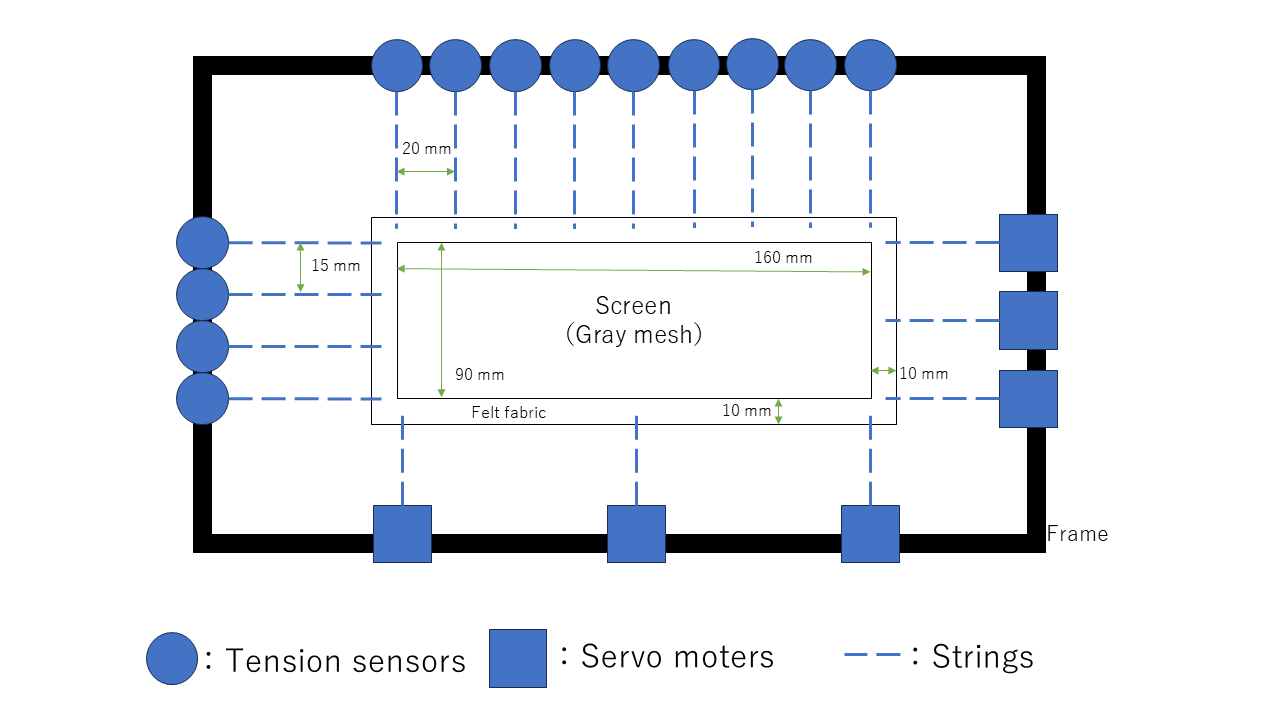}\\
    (b) Sensor and motor layout
  \end{minipage}\hfill
  \begin{minipage}[b]{0.18\textwidth}
    \centering\includegraphics[width=\linewidth]{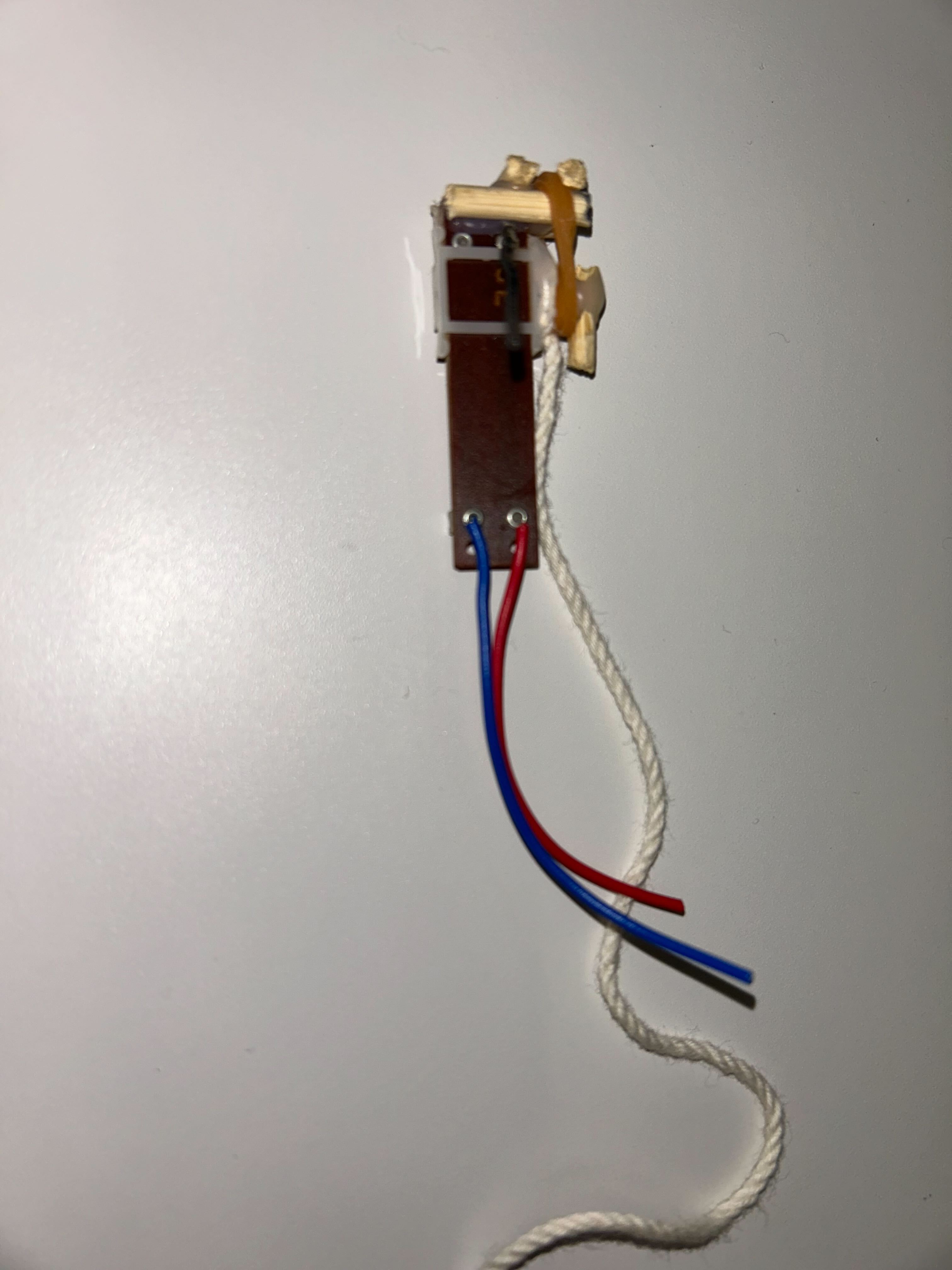}\\
    (c) Tension sensor
  \end{minipage}
  \Description{Photograph of the DAT mesh and projector, schematic showing sensors and servomotors around the screen, and a close-up of the sliding-potentiometer tension sensor.}
  \caption{DAT screen, hardware layout, and tension sensor. Adapted from \cite{Tamamoto2025DAT}.}
  \label{fig:dat-system}
\end{figure*}

\section{Introduction}

Virtual agents commonly communicate through appearance, movement, and speech. Touch offers another possible way to interact with them: users can reach toward and make contact with an agent's apparent body. Yet making such contact possible also raises questions about how the agent should respond and how users may interpret that response. The same touch input could be treated as a command or as an action toward the agent.

We built DAT on the Amid Screen, a low-cost technique for spatial projection using semi-transparent window-screen mesh \cite{AokiAmidScreen}. The mesh serves as a projection surface while remaining flexible enough to be pressed. DAT extends this surface with touch sensing: rather than adding a rigid touch layer, it detects changes in the tension of the threads supporting the mesh. The first prototype estimated the location and depth of one press \cite{Tamamoto2024AmidTouch}; the second added servomotors that vary the resistance encountered when pressing the screen \cite{Tamamoto2025DAT}. Taken together, the prototypes provide a platform for projecting an agent onto a compliant surface, sensing contact with that surface, and changing its mechanical response.

Our interest here extends beyond any one demonstration. When a displayed agent responds to contact, designers can make its reaction appear to express a preference or a boundary. This possibility motivates a question for the workshop: should systems such as DAT implement agents that appear to have such agency, and what would justify that choice? We first describe the technological basis and the existing demonstrations so that the question is grounded in what the system can actually do. We do not infer the user's intention, the agent's consent, or an effect on social perception from press depth alone.

\section{Background and Related Work}
 \subsection{Spatial projection and touch input}
Spatial projection has been realized with transparent projection surfaces such as agricultural polyethylene film in Polid Screen \cite{PolidScreen} and window-applied rear-projection film in DILAD Screen \cite{KimotoDiladScreen}, as well as with a flowing mist layer \cite{Kume2002MistProjection}. The Amid Screen uses inexpensive, partly transparent window-screen mesh as its projection surface \cite{AokiAmidScreen}. We built DAT on this mesh because it supports spatial projection and can be suspended by threads whose tension changes when the mesh is pressed.

Conventional sensing methods solve related input problems on different surfaces. Koyama et al. used a depth sensor at the side of a transparent screen to detect touch from either side \cite{Koyama2014DoubleSided}. Han's frustrated total internal reflection technique supports multiple touches on a rigid optical surface \cite{Han2005FTIR}. These examples establish ways to sense contact with transparent displays; the DAT screen instead exploits changes in the threads that suspend a flexible projection surface. The resulting signal can be used to estimate where a single press occurs and how deeply the mesh is pushed \cite{Tamamoto2024AmidTouch}.

\subsection{Changing the response to a press}
Touchscreen feedback can be delivered through vibration, as in the display studied by Mishima et al. \cite{Mishima2015Vibration}. Other interactive displays can change surface shape and stiffness \cite{Ikeda2016VariableDisplay}. The DAT screen approach adjusts tension in the threads supporting a projected mesh \cite{Tamamoto2025DAT}. Here, \emph{virtual hardness} denotes a change in resistance to indentation produced by that tension control; it does not mean that the intrinsic material hardness of a displayed body has been measured. This distinction matters when connecting a mechanical interface to the apparent behavior of a virtual agent.

\begin{figure}[t]
  \centering
  \includegraphics[width=\columnwidth]{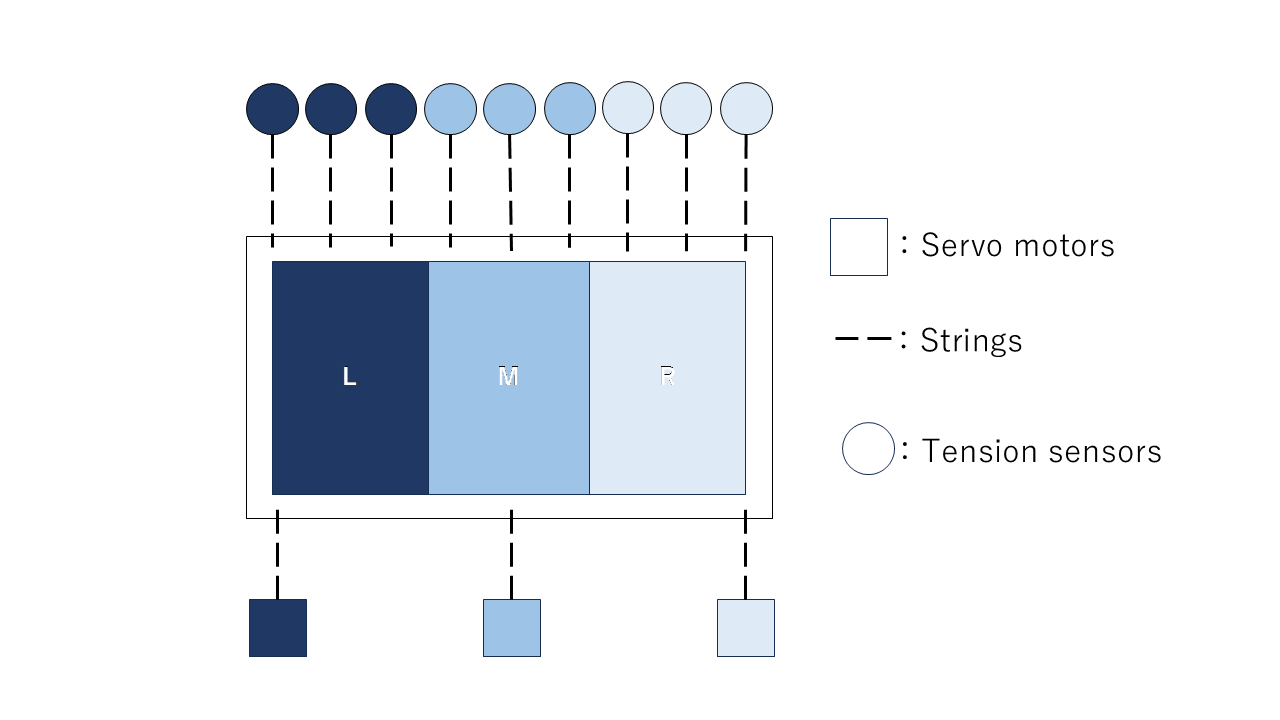}
  \Description{Diagram dividing the DAT screen into left, center, and right regions and showing the corresponding servomotors and tension sensors.}
  \caption{Screen regions for tension control. Adapted from \cite{Tamamoto2025DAT}.}
  \label{fig:region-control}
\end{figure}

\begin{figure*}[t]
  \centering
  \begin{minipage}[t]{0.48\textwidth}
    \centering
    \includegraphics[width=\linewidth]{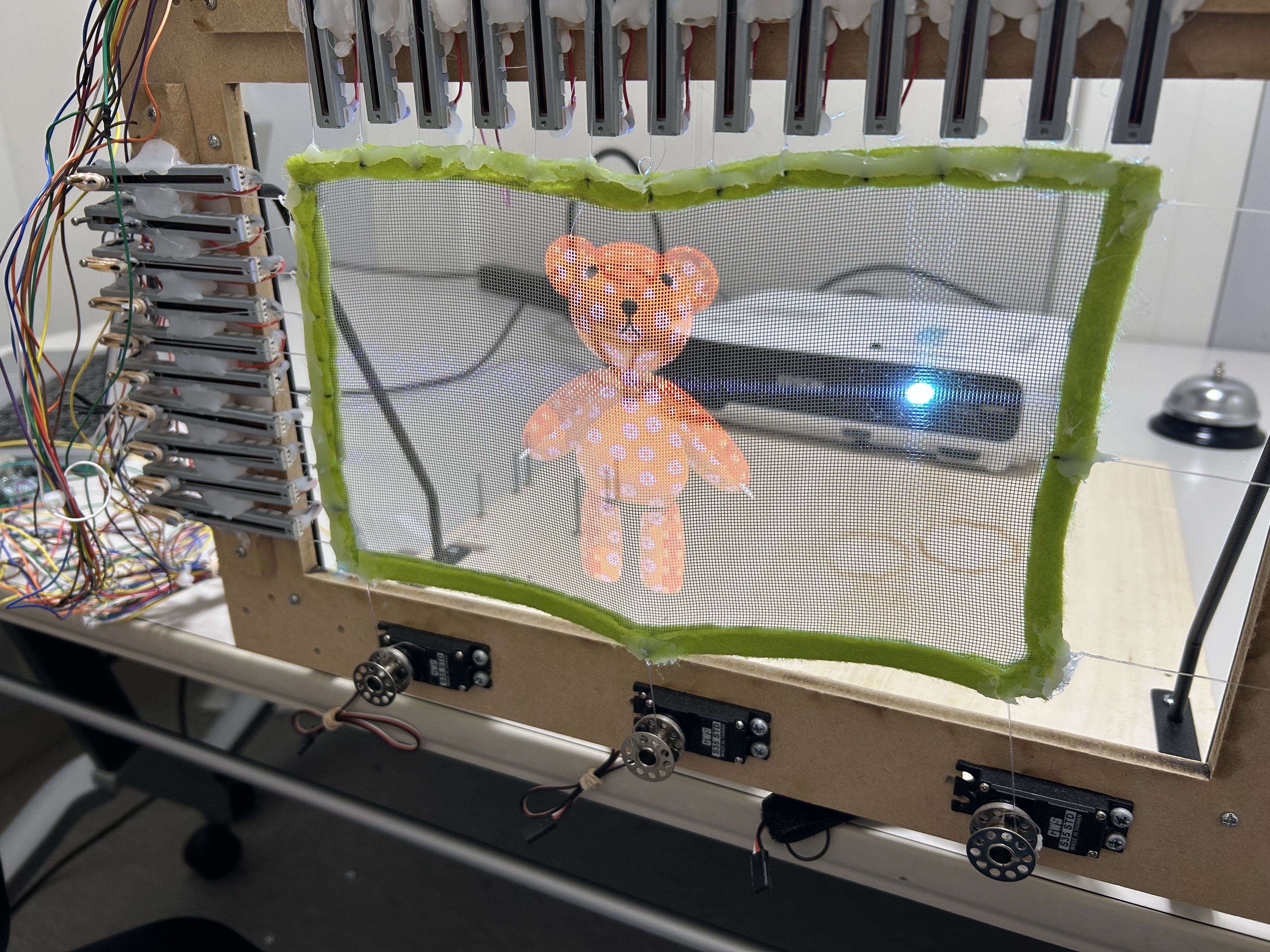}
    \par\smallskip
    {\small (a) Default state}
  \end{minipage}\hfill
  \begin{minipage}[t]{0.48\textwidth}
    \centering
    \includegraphics[width=\linewidth]{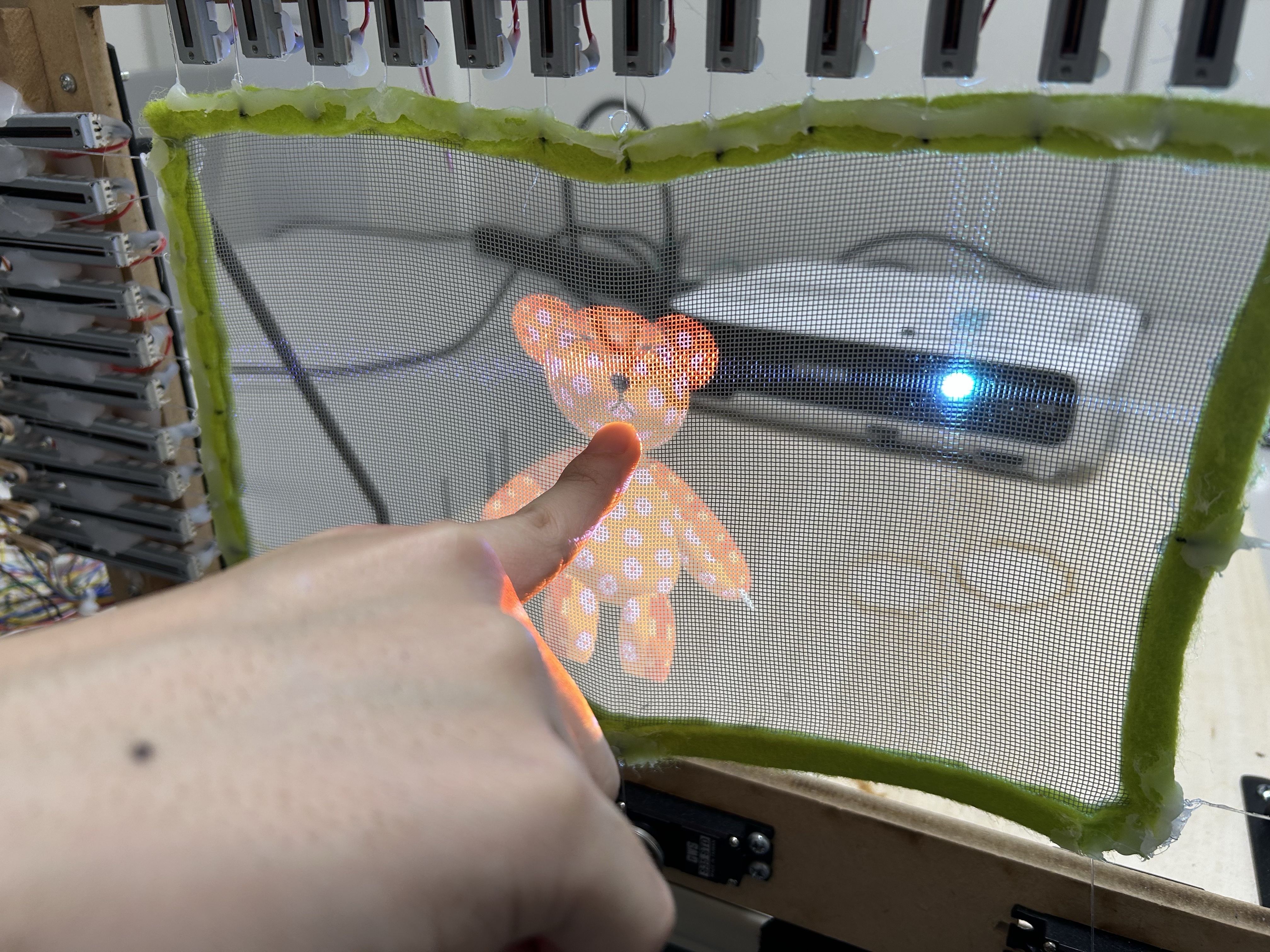}
    \par\smallskip
    {\small (b) Seated after a fall, with eyes closed}
  \end{minipage}

  \caption{On-site demonstration of DAT.}
  \Description{Two views of the projected agent: its default standing
  state and its seated state after a fall, with its eyes closed.}
  \label{fig:agent-fall-demo}

  \par\noindent{\footnotesize
  \emph{Model credit:} \emph{Usagi to Kuma no Nuigurumi}
  by atuyu (@atuyu\_no01), modified by the authors
  (additional eye variants);
  \url{https://bowlroll.net/file/146474}.\par}
\end{figure*}

\section{The DAT Screen System}

\subsection{Projection and Touch Sensing}
The DAT screen combines a projector, a semi-transparent gray window-screen mesh, and threads connecting the mesh to tension sensors. Images appear on the mesh while the projection surface remains physically pressable. Tension-based single-touch sensing was first reported in the 2024 DAT implementation \cite{Tamamoto2024AmidTouch}. Here, we focus on the tension-controlled implementation, which adds servomotors to vary the mesh's resistance to indentation \cite{Tamamoto2025DAT}. Figure~\ref{fig:dat-system} shows the device, its sensor and motor layout, and a close-up of a tension sensor.

The tension-controlled screen has a $90\,\mathrm{mm} \times 160\,\mathrm{mm}$ projection area and 13 tension sensors arranged along two perpendicular axes. Each sensor uses a thread attached to a sliding potentiometer: pressing the mesh changes the tension in nearby threads and thus the sensor readings. Touch position is estimated by applying cubic-spline interpolation to the changes along each axis, while the sum of those changes serves as an indicator of indentation depth \cite{Tamamoto2024AmidTouch,Tamamoto2025DAT}. Motor actuation also changes the sensor readings, so touch-induced changes must be interpreted in relation to the current tension state.

\subsection{Controlling Resistance to Indentation}
Servomotors wind or release threads attached to the mesh. For screen-wide control, the motors operate until the sum of the sensor readings reaches one of two target ranges, defining an initial state and a higher-tension state \cite{Tamamoto2025DAT}. In a mechanism check, presses to the same depth of $15\,\mathrm{mm}$ produced larger sensor changes in the higher-tension state. The reported result supports a difference in mechanical response; perceived hardness was not evaluated in a controlled study. The motors can also be controlled by screen region, allowing resistance to be varied across left, center, and right regions rather than independently at every point. Figure~\ref{fig:region-control} shows the correspondence among screen regions, motors, and sensors.

\subsection{Demonstrations and Existing Evidence}
The published tension-control demonstration paired drawings of a relaxed and a contracted arm muscle with the two tension states \cite{Tamamoto2025DAT}. A separate on-site demonstration linked touch position and press depth to an agent's reactions: a shallow press could elicit a positive response to being stroked, whereas a deeper press could make the agent fall, get back up, turn away, and appear to cry. Figure~\ref{fig:agent-fall-demo} shows the press and the resulting fall in this demonstration. This illustrates a possible agent interaction, but it has not established that visitors interpreted the latter reaction as refusal or as a boundary.

A separate study found higher ratings for the sense of touching a living thing and for some measures of an agent's presence with a more deformable surface, while an improvement due to transparency was not confirmed \cite{Tamamoto2026Presence}. It did not examine whether users interpret an agent's response to a deep press as refusal.






\section{Discussion}
\subsection{Possible Interaction and Development}
The DAT screen's tension control and spatial projection suggest possible ways to express an agent's reluctance toward contact. 

For example, increased resistance to indentation, paired with an agent's fearful movement, might evoke the impression of a body stiffening. Spatial projection could instead show the agent withdrawing from the projected contact area, leaving the user touching only the semi-transparent mesh. These are design possibilities rather than interactions evaluated in the present work; whether users would interpret them as fear or avoidance remains an open question.
%
\subsection{Reality, Presence, Trust, and Blame of Agency}
An agent that appears present and responds to touch may lead users to read a withdrawal, a fearful animation, or increased resistance as an expression of its own preference. However, DAT detects properties of a press and presents responses chosen by its designers; it cannot infer the user's intention or establish the agent's consent. The reported presence ratings \cite{Tamamoto2026Presence} do not tell us whether users would trust such a response as a meaningful refusal. Such a refusal cue might also impose an unintended emotional burden on the user. Responsibility for that possibility and for what the response communicates rests with the designers, rather than with the displayed agent. The question of boundaries also extends beyond how a user touches the agent.

\paragraph{Cross-boundary Interaction}
A user might deliberately continue pressing after an agent appears to withdraw or object, perhaps to elicit a fearful reaction. On DAT, a visual withdrawal would not by itself end physical contact: the user's finger could continue deforming the mesh and feeling its resistance. How should a DAT-like system respond to such potentially abusive behavior: by repeating the objection, making the agent withdraw further, or ending the interaction? The effects of these choices on users have not been evaluated.

Boundaries can also be crossed in the other direction: a projected agent might approach the user unexpectedly, or the screen might change its resistance during contact without the user requesting that change. These would be visual or tactile interactions, not physical contact initiated by the agent. 

Whether such reciprocal interactions are desirable, and how users should be able to stop them, requires investigation.

\subsection{Limitation and Future Perspective}
Taken together, our DAT demonstrations show that the screen can detect presses, vary resistance, and link touch input to an agent's visual response. We have not established whether users interpret that response as a preference or refusal, or how they would react if contact continued after an apparent refusal. Our separate presence study used different stimuli and did not address these questions \cite{Tamamoto2026Presence}.

Future work should examine how users interpret visual withdrawal and changes in resistance, both separately and together. It should also investigate agent-initiated interactions and whether users can recognize and stop an interaction they do not want. These findings would help determine whether and how boundary-like behavior should be implemented in DAT-like systems.

\begin{acks}
This research was supported in part by JSPS KAKENHI Grant Numbers 23K11278, 26K14911, 23K11202, and 24K02977. The discussion of safety, user-agent boundaries, and responsibility for potentially unwanted interactions was supported by the JST Moonshot R\&D Program (Grant Number JPMJMS2215).
\end{acks}

\bibliographystyle{ACM-Reference-Format}
 \bibliography{bunken}
\end{document}